\documentclass[journal]{IEEEtran}
\usepackage{amsmath,amssymb,amsfonts}
\usepackage{algorithmic}
\usepackage{graphicx}
\usepackage{hyperref}
\usepackage{textcomp}
\usepackage{xcolor}
\usepackage{tikz}
\usetikzlibrary{shapes.geometric, arrows, positioning}
\usepackage{cite}

\def\BibTeX{{\rm B\kern-.05em{\sc i\kern-.025em b}\kern-.08em
    T\kern-.1667em\lower.7ex\hbox{E}\kern-.125emX}}

\begin{document}

\title{Automated Statistical and Machine Learning Platform for Biological Research}

\title{Automated Statistical and Machine Learning Platform for Biological Research}

\author{Luke~Rimmo~Lego\IEEEauthorrefmark{1}, Samantha~Gauthier\IEEEauthorrefmark{2}, and Denver~Jn.~Baptiste\IEEEauthorrefmark{3}%
\thanks{\IEEEauthorrefmark{1}Luke Rimmo Lego, Student, Department of Biomedical Engineering, Stevens Institute of Technology, Hoboken, NJ 07030, USA.}%
\thanks{\IEEEauthorrefmark{2}Samantha Gauthier, Student, Department of Computer Science, Stevens Institute of Technology, Hoboken, NJ 07030, USA.}%
\thanks{\IEEEauthorrefmark{3}Denver Jn. Baptiste, Principal Investigator, Department of Chemistry and Chemical Biology, Stevens Institute of Technology, Hoboken, NJ 07030, USA. Corresponding author: \texttt{djnbaptiste@stevens.edu}.}%
}

\maketitle

\begin{abstract}
Research increasingly relies on computational methods to analyze experimental data and predict molecular properties. Current approaches often require researchers to use a variety of tools for statistical analysis and machine learning, creating workflow inefficiencies. We present an integrated platform that combines classical statistical methods with Random Forest classification for comprehensive data analysis that can be used in the biological sciences. The platform implements automated hyperparameter optimization, feature importance analysis, and a suite of statistical tests including t-tests, ANOVA, and Pearson correlation analysis. Our methodology addresses the gap between traditional statistical software, modern machine learning frameworks and biology, by providing a unified interface accessible to researchers without extensive programming experience. The system achieves this through automatic data preprocessing, categorical encoding, and adaptive model configuration based on dataset characteristics. Initial testing protocols are designed to evaluate classification accuracy across diverse chemical datasets with varying feature distributions. This work demonstrates that integrating statistical rigor with machine learning interpretability can accelerate biological discovery workflows while maintaining methodological soundness. The platform’s modular architecture enables future extensions to additional machine learning algorithms and statistical procedures relevant to bioinformatics.
\end{abstract}


\section{Introduction}

Modern biological research generates vast quantities of experimental data that require sophisticated analytical methods to extract meaningful insights. Researchers must often navigate between statistical software packages for hypothesis testing and separate machine learning frameworks for predictive mod- eling. This fragmentation creates barriers to reproducibility and slows the research process, particularly for investigators without formal training in computer science or data science methodologies. 

The challenge of integrating multiple analytical paradigms becomes particularly acute in biology, where datasets often contain mixed variable types including continuous mea- surements such as spectroscopic data alongside categorical features like biological classifications or reaction conditions. Traditional statistical software excels at hypothesis testing but provides limited support for predictive modeling, while machine learning libraries currently lack comprehensive statistical testing capabilities. This dichotomy forces researchers to 

export data between tools, increasing the risk of transcription errors and complicating workflow documentation. 

Recent advances in browser-based computation and JavaScript machine learning libraries have created opportunities to deliver integrated analytical platforms without re- quiring local software installation or specialized computing infrastructure. However, existing web-based tools tend to focus narrowly on either statistical analysis or machine learning, rarely combining both with appropriate emphasis on interpretability and statistical rigor. Furthermore, many tools lack automated optimization features that would make them accessible to researchers unfamiliar with hyperparameter tuning or model selection procedures. 

This work addresses these limitations by implementing an integrated platform that seamlessly combines exploratory data analysis, classical hypothesis testing, and Random Forest classification within a unified interface. The system automates key decisions regarding data preprocessing, model configuration, and feature importance calculation while maintaining transparency about its analytical choices. Our approach prioritizes interpretability and statistical soundness over pure predictive performance, recognizing that chemical researchers require not only accurate predictions but also mechanistic understanding of the factors driving those predictions.

\section{Background}

\subsection{Random Forest Classification}

Random Forest represents an ensemble learning method that constructs multiple decision trees during training and outputs the mode of their individual predictions for classification tasks \cite{breiman2001random}. The algorithm introduces randomness through bootstrap aggregating of training samples and random feature selection at each node split. This dual randomization strategy reduces overfitting while maintaining high predictive accuracy across diverse problem domains \cite{hastie2009elements}.

The theoretical foundation of Random Forest relies on the strength of individual trees and the correlation between them. Strong trees with low correlation produce more accurate ensemble predictions. For a Random Forest classifier with trees $h_1(\mathbf{x}), h_2(\mathbf{x}), \ldots, h_K(\mathbf{x})$, the margin function for input $\mathbf{x}$ with true class $y$ is defined as:

\begin{equation}
mg(\mathbf{x}, y) = P_\Theta(h(\mathbf{x}, \Theta) = y) - \max_{j \neq y} P_\Theta(h(\mathbf{x}, \Theta) = j)
\end{equation}

where $\Theta$ represents the random parameter vector determining tree construction.  This allows the model to select the correct answers over the error prone ones.  The generalization error is then bounded by:

\begin{equation}
PE^* \leq \frac{\bar{\rho}(1-s^2)}{s^2}
\end{equation}

where $s$ is the strength of the ensemble and $\bar{\rho}$ is the mean correlation between trees \cite{breiman2001random}. This relationship demonstrates that increasing tree strength while decreasing inter-tree correlation improves classification performance, motivating the random subspace method employed at each split.

\subsection{Feature Importance Metrics}

Random Forest provides intrinsic measures of feature importance through permutation-based assessment and mean decrease in impurity. The permutation importance for feature $j$ quantifies the increase in prediction error when that feature's values are randomly shuffled, breaking its relationship with the target variable. Formally, for out-of-bag samples:

\begin{equation}
I_j = \frac{1}{K} \sum_{k=1}^{K} \left( E_k^{perm,j} - E_k^{orig} \right)
\end{equation}

where $E_k^{orig}$ is the original out-of-bag error for tree $k$ and $E_k^{perm,j}$ is the error after permuting feature $j$ \cite{strobl2007bias}. This metric reveals which features contribute most substantially to predictive accuracy, providing interpretable insights into the underlying data structure.

\subsection{Statistical Hypothesis Testing}

Classical hypothesis testing frameworks remain essential for establishing statistical significance in chemical research. The independent samples t-test evaluates whether two population means differ significantly, with the test statistic:

\begin{equation}
t = \frac{\bar{X}_1 - \bar{X}_2}{\sqrt{\frac{s_1^2}{n_1} + \frac{s_2^2}{n_2}}}
\end{equation}

where $\bar{X}_i$, $s_i^2$, and $n_i$ represent the sample mean, variance, and size for group $i$ \cite{student1908probable}. Under the null hypothesis of equal means, this statistic follows a t-distribution with degrees of freedom approximated by the Welch-Satterthwaite equation.

For comparing multiple groups, analysis of variance (ANOVA) partitions total variance into between-group and within-group components. The F-statistic:

\begin{equation}
F = \frac{\text{MS}_{between}}{\text{MS}_{within}} = \frac{\sum_{i=1}^{k} n_i(\bar{X}_i - \bar{X})^2 / (k-1)}{\sum_{i=1}^{k}\sum_{j=1}^{n_i}(X_{ij} - \bar{X}_i)^2 / (N-k)}
\end{equation}

tests the null hypothesis that all group means are equal \cite{fisher1925statistical}. Significant F-statistics require post-hoc testing to identify which specific groups differ, for which we employ Tukey's Honest Significant Difference method.

\subsection{Correlation Analysis}

Pearson correlation quantifies linear relationships between continuous variables. For paired observations $(x_i, y_i)$, the correlation coefficient is:

\begin{equation}
r = \frac{\sum_{i=1}^{n}(x_i - \bar{x})(y_i - \bar{y})}{\sqrt{\sum_{i=1}^{n}(x_i - \bar{x})^2 \sum_{i=1}^{n}(y_i - \bar{y})^2}}
\end{equation}

This coefficient ranges from -1 to 1, where values near zero indicate weak linear association \cite{pearson1895correlation}. Statistical significance testing transforms $r$ to a t-statistic with $n-2$ degrees of freedom, enabling hypothesis tests regarding correlation strength.

\section{Methodology}

\subsection{System Architecture}

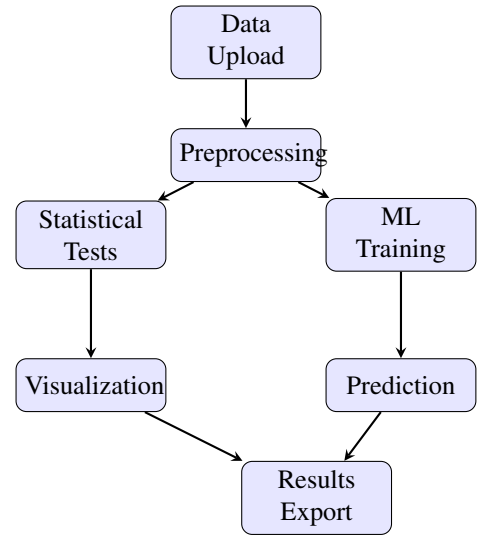
\begin{figure}[h]
\centering
\begin{tikzpicture}[node distance=1.5cm, auto,
    block/.style={rectangle, draw, fill=blue!10, text width=5em, text centered, rounded corners, minimum height=2em},
    arrow/.style={thick,->,>=stealth}]

    \node [block] (upload) {Data Upload};
    \node [block, below of=upload] (preprocess) {Preprocessing};
    \node [block, below left of=preprocess, xshift=-1cm] (stats) {Statistical Tests};
    \node [block, below right of=preprocess, xshift=1cm] (ml) {ML Training};
    \node [block, below of=stats, yshift=-0.5cm] (viz1) {Visualization};
    \node [block, below of=ml, yshift=-0.5cm] (pred) {Prediction};
    \node [block, below of=viz1, xshift=3cm] (export) {Results Export};

    \draw [arrow] (upload) -- (preprocess);
    \draw [arrow] (preprocess) -- (stats);
    \draw [arrow] (preprocess) -- (ml);
    \draw [arrow] (stats) -- (viz1);
    \draw [arrow] (ml) -- (pred);
    \draw [arrow] (viz1) -- (export);
    \draw [arrow] (pred) -- (export);
\end{tikzpicture}
\caption{System workflow demonstrating data flow from upload through analysis to results.}
\label{fig:workflow}
\end{figure}

The platform architecture follows a modular design separating data ingestion, preprocessing, analysis, and visualization components. Users upload CSV-formatted datasets through a browser interface, which parses the data using the PapaParse library. The system automatically infers column types by examining data distributions, classifying features as numeric or categorical based on unique value counts and data type consistency.

\subsection{Data Preprocessing Pipeline}

Categorical features undergo label encoding, mapping string values to integer indices. For a categorical feature $x$ with unique values $\{v_1, v_2, \ldots, v_m\}$, we define the encoding function:

\begin{equation}
f(x) = \begin{cases}
0 & \text{if } x = v_1 \\
1 & \text{if } x = v_2 \\
\vdots \\
m-1 & \text{if } x = v_m
\end{cases}
\end{equation}

This preserves categorical information while enabling numerical computation. We maintain encoding dictionaries to ensure consistent transformations between training and prediction phases.

Numeric features undergo z-score normalization to prevent scale differences from biasing model training. For each numeric feature $X$, we compute:

\begin{equation}
Z = \frac{X - \mu}{\sigma}
\end{equation}

where $\mu$ and $\sigma$ represent the feature's mean and standard deviation calculated from training data. This transformation centers features at zero with unit variance, satisfying the standardization assumptions of many statistical procedures.

\subsection{Automated Hyperparameter Optimization}

Rather than requiring manual hyperparameter specification, the system implements adaptive optimization based on dataset characteristics. The number of trees $K$ scales with sample size $n$ and feature count $p$:

\begin{equation}
K = \min\left(\left\lceil 50 + 10 \log(n) \right\rceil, 200\right)
\end{equation}

This formula ensures sufficient ensemble diversity for small datasets while preventing excessive computation for large samples. Maximum tree depth $d_{max}$ adapts to feature dimensionality:

\begin{equation}
d_{max} = \min\left(\left\lceil 5 + \sqrt{p} \right\rceil, 20\right)
\end{equation}

preventing overfitting in high-dimensional spaces while allowing sufficient model complexity for feature interactions. Minimum samples per leaf node follows:

\begin{equation}
s_{min} = \max\left(\left\lfloor 0.01n \right\rfloor, 2\right)
\end{equation}

ensuring leaf nodes contain sufficient samples for reliable probability estimates.

\subsection{Train-Test Splitting}

We partition datasets using stratified random sampling with an 80-20 train-test split. For a dataset with $N$ samples distributed across $C$ classes with proportions $p_1, p_2, \ldots, p_C$, stratification ensures:

\begin{equation}
\left|\frac{n_{train,c}}{N_{train}} - \frac{n_{test,c}}{N_{test}}\right| < \epsilon
\end{equation}

for all classes $c$, where $\epsilon$ represents a small tolerance value. This preserves class distribution across splits, preventing evaluation bias from unbalanced partitions.

\subsection{Performance Metric Calculation}

We evaluate model performance using multiple complementary metrics. Classification accuracy provides overall correctness:

\begin{equation}
\text{Accuracy} = \frac{TP + TN}{TP + TN + FP + FN}
\end{equation}

where $TP$, $TN$, $FP$, and $FN$ denote true positives, true negatives, false positives, and false negatives. For imbalanced datasets, precision and recall offer additional perspective:

\begin{equation}
\text{Precision} = \frac{TP}{TP + FP}, \quad \text{Recall} = \frac{TP}{TP + FN}
\end{equation}

The F1-score harmonically combines these metrics:

\begin{equation}
F_1 = 2 \cdot \frac{\text{Precision} \cdot \text{Recall}}{\text{Precision} + \text{Recall}}
\end{equation}

providing a balanced assessment of classification quality. For probabilistic predictions, we compute receiver operating characteristic curves by varying decision thresholds $\tau$ and plotting true positive rate against false positive rate:

\begin{equation}
TPR(\tau) = \frac{|\{i: \hat{p}_i \geq \tau \land y_i = 1\}|}{|\{i: y_i = 1\}|}
\end{equation}

\begin{equation}
FPR(\tau) = \frac{|\{i: \hat{p}_i \geq \tau \land y_i = 0\}|}{|\{i: y_i = 0\}|}
\end{equation}

The area under this curve (AUC) summarizes classifier discrimination ability across all possible thresholds \cite{hanley1982meaning}.

\subsection{Statistical Testing Implementation}

For t-tests, we compute the test statistic using pooled or unpooled variance estimates depending on homogeneity assumptions. The two-tailed p-value derives from the cumulative t-distribution:

\begin{equation}
p = 2P(T > |t_{obs}|)
\end{equation}

where $T$ follows the appropriate t-distribution. ANOVA employs the F-distribution for p-value calculation, with post-hoc Tukey tests computing all pairwise comparisons using the studentized range distribution.

Pearson correlation testing transforms the correlation coefficient to a t-statistic:

\begin{equation}
t = r\sqrt{\frac{n-2}{1-r^2}}
\end{equation}

which follows a t-distribution with $n-2$ degrees of freedom under the null hypothesis of zero correlation \cite{student1908probable}.

\section{Theory and Analysis}

\subsection{Random Forest Convergence Properties}

The law of large numbers guarantees that Random Forest predictions converge as the number of trees increases. Let $h_1, h_2, \ldots, h_K$ represent independent identically distributed trees with individual error rate $e < 0.5$. The ensemble error rate $E_K$ satisfies:

\begin{equation}
\lim_{K \to \infty} E_K = e
\end{equation}

with probability 1. Furthermore, for finite $K$, the variance of the ensemble prediction decreases as:

\begin{equation}
\text{Var}(\bar{h}) = \frac{\sigma^2}{K}
\end{equation}

where $\sigma^2$ represents the variance of individual tree predictions \cite{breiman2001random}. This theoretical foundation justifies our adaptive selection of tree counts based on dataset size.

\subsection{Bias-Variance Tradeoff in Tree Depth}

Tree depth directly controls model complexity and influences the bias-variance tradeoff. Deep trees exhibit low bias but high variance, memorizing training data and failing to generalize. Shallow trees maintain high bias but low variance, underfitting complex patterns. The expected prediction error decomposes as:

\begin{equation}
E[(y - \hat{f}(x))^2] = \text{Bias}[\hat{f}(x)]^2 + \text{Var}[\hat{f}(x)] + \sigma^2
\end{equation}

where $\sigma^2$ represents irreducible error. Random Forest partially mitigates this tradeoff through ensemble averaging, which reduces variance while maintaining low bias from individual deep trees \cite{hastie2009elements}.

\subsection{Feature Importance Validity}

Permutation importance provides unbiased feature ranking under the assumption that features are mutually independent. When features exhibit correlation, permutation may create unrealistic data combinations, potentially underestimating importance for correlated predictors. Consider features $X_1$ and $X_2$ with correlation $\rho$. Permuting $X_1$ generates pairs $(X_1^{perm}, X_2)$ with correlation:

\begin{equation}
\text{Cor}(X_1^{perm}, X_2) \approx 0
\end{equation}

This decorrelation may inflate error estimates if both features jointly predict the target. Despite this limitation, permutation importance remains interpretable and computationally efficient for preliminary feature screening \cite{strobl2007bias}.

\subsection{Multiple Comparison Corrections}

Statistical testing across multiple hypotheses requires correction for family-wise error rate inflation. Without adjustment, the probability of at least one Type I error grows with the number of tests $m$:

\begin{equation}
P(\text{at least one Type I error}) = 1 - (1-\alpha)^m
\end{equation}

For Tukey's HSD procedure, we control this rate by comparing pairwise differences to critical values from the studentized range distribution:

\begin{equation}
q = \frac{|\bar{X}_i - \bar{X}_j|}{\sqrt{\text{MSE}/n}}
\end{equation}

where MSE denotes the mean squared error from ANOVA. This method maintains the family-wise error rate at the nominal $\alpha$ level across all comparisons \cite{tukey1949comparing}.

\section{System Design and Implementation}

\subsection{User Interface Architecture}

The platform implements a tab-based navigation structure organizing functionality into three primary modules: Dashboard, AI Training, and Data Analysis. This separation reflects distinct user workflows while maintaining visual consistency through a unified design system built on React components and Tailwind CSS. The interface prioritizes progressive disclosure, presenting essential controls prominently while making advanced options accessible through expandable sections.

The Dashboard serves as the entry point, providing quick access to recent analyses and model versions. Visual cards display key metadata including dataset characteristics, model performance metrics, and creation timestamps. Users navigate between modules through a persistent header containing the institution logo and page links, establishing clear spatial orientation throughout the application.

\subsection{Data Upload and Parsing}

Data ingestion begins with drag-and-drop CSV file upload implemented through a browser-based file reader. The PapaParse library handles CSV parsing with automatic delimiter detection and header inference. Upon upload, the system performs immediate validation checking for empty files, inconsistent column counts, and unparseable content. Error messages provide specific guidance when validation fails, directing users to correct formatting issues.

The parser constructs a data matrix representation where rows correspond to observations and columns to features. String values undergo trimming to remove whitespace, and empty cells receive special handling depending on column type. The system maintains the original column ordering from the uploaded file, preserving any semantic arrangement researchers may have established.

\subsection{Interactive Data Preview}

Following successful upload, a scrollable table displays the first 100 rows of data with fixed headers. Column names appear prominently, and cells render with alternating row colors improving visual scanning. Numeric columns align right while text columns align left, following conventional tabular data presentation. Users can verify data integrity before proceeding to analysis, catching upload errors or formatting problems early in the workflow.

Summary statistics appear below the preview table, including total row count, column count, and per-column data type classification. The system distinguishes numeric columns containing only numbers from categorical columns with string or mixed content. This automatic type inference eliminates manual schema specification while remaining transparent about its decisions.

\subsection{Model Training Interface}

The AI Training module presents a streamlined interface for Random Forest configuration. Users select a target column from a dropdown menu populated with all categorical columns detected in their dataset. Feature selection occurs automatically, with the system including all columns except the target in the feature matrix. This default behavior works well for curated datasets where irrelevant features have been removed during data preparation.

A single "Train Model" button initiates the training process, which executes asynchronously to maintain interface responsiveness. During training, a progress indicator displays the current stage: data splitting, preprocessing, model fitting, or evaluation. This feedback reduces uncertainty during lengthy computations on large datasets.

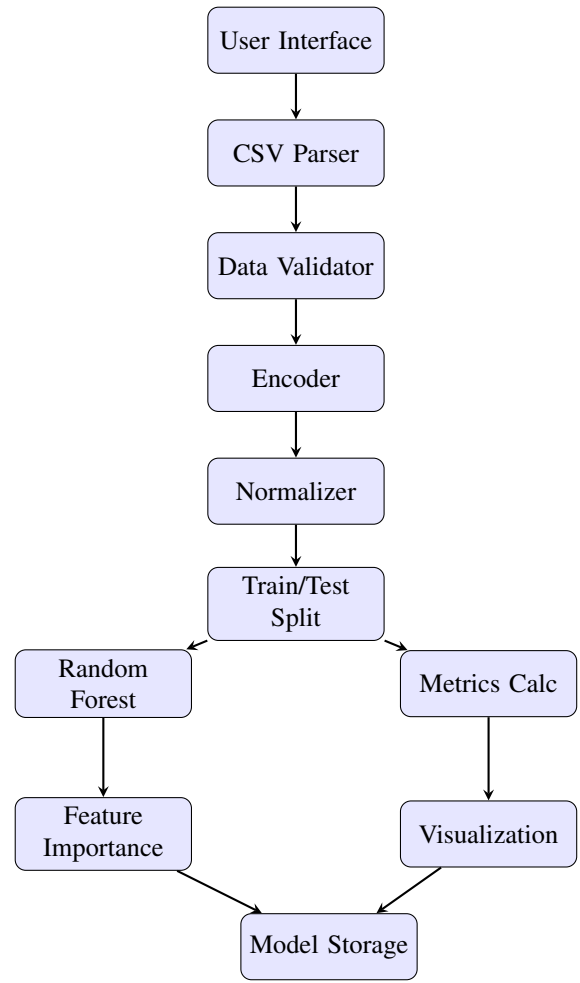
\begin{figure}[h]
\centering
\begin{tikzpicture}[node distance=1.5cm, auto,
    block/.style={rectangle, draw, fill=blue!10, text width=6em, text centered, rounded corners, minimum height=2.5em},
    arrow/.style={thick,->,>=stealth}]

    \node [block] (ui) {User Interface};
    \node [block, below of=ui] (parse) {CSV Parser};
    \node [block, below of=parse] (validate) {Data Validator};
    \node [block, below of=validate] (encode) {Encoder};
    \node [block, below of=encode] (normalize) {Normalizer};
    \node [block, below of=normalize] (split) {Train/Test Split};
    \node [block, below left of=split, xshift=-1.5cm] (rf) {Random Forest};
    \node [block, below right of=split, xshift=1.5cm] (metrics) {Metrics Calc};
    \node [block, below of=rf, yshift=-0.5cm] (importance) {Feature Importance};
    \node [block, below of=metrics, yshift=-0.5cm] (viz) {Visualization};
    \node [block, below of=importance, xshift=3cm] (storage) {Model Storage};

    \draw [arrow] (ui) -- (parse);
    \draw [arrow] (parse) -- (validate);
    \draw [arrow] (validate) -- (encode);
    \draw [arrow] (encode) -- (normalize);
    \draw [arrow] (normalize) -- (split);
    \draw [arrow] (split) -- (rf);
    \draw [arrow] (split) -- (metrics);
    \draw [arrow] (rf) -- (importance);
    \draw [arrow] (metrics) -- (viz);
    \draw [arrow] (importance) -- (storage);
    \draw [arrow] (viz) -- (storage);
\end{tikzpicture}
\caption{Detailed implementation flow from user interaction through model storage.}
\label{fig:implementation}
\end{figure}

\subsection{Results Visualization Components}

Training completion triggers display of multiple coordinated visualizations. A confusion matrix heatmap uses color intensity to encode cell counts, with darker shades indicating higher frequencies. Row and column labels identify predicted and actual classes respectively. This representation makes classification errors immediately visible through off-diagonal elements.

The ROC curve plots as a line chart with false positive rate on the x-axis and true positive rate on the y-axis. A diagonal reference line representing random chance provides context for evaluating classifier discrimination. The area under the curve appears as a numeric annotation, enabling quick performance assessment. Interactive tooltips display exact coordinates when users hover over the curve.

Feature importance renders as a horizontal bar chart sorted by importance magnitude. Feature names label the y-axis while bar lengths encode importance values on the x-axis. This layout accommodates long feature names better than vertical bars would. Color coding distinguishes highly important features from less influential ones, drawing attention to key predictors.

\subsection{Model Version Management}

The system maintains a history of trained models in browser local storage, enabling comparison across different target variables or dataset versions. Each saved model includes the complete feature preprocessing pipeline, tree ensemble parameters, and evaluation metrics. A table-based interface lists models chronologically with columns for creation date, target variable, accuracy, and feature count.

Users can load previously trained models for making predictions on new data without retraining. This functionality proves valuable when applying validated models to additional samples or when comparing predictions across model versions. The storage implementation uses JSON serialization for model parameters and encoding dictionaries, maintaining full reproducibility.

\subsection{Prediction Workflow}

After training or loading a model, users access the prediction interface through a dedicated form. Input fields appear for each feature in the training dataset, with field types matching the original data types. Categorical features present dropdown menus populated with values observed during training, preventing invalid inputs. Numeric features accept free text input with validation ensuring numeric format.

Upon form submission, the system applies the saved preprocessing transformations before invoking the model. Categorical values map to their training-phase encodings, and numeric values undergo z-score normalization using stored means and standard deviations. The model then generates class probabilities for each possible outcome. The interface displays the predicted class prominently alongside probability distributions for all classes, giving users insight into prediction confidence.

\subsection{Statistical Analysis Interface}

The Data Analysis module provides separate cards for each statistical procedure: descriptive statistics, t-tests, ANOVA, and correlation analysis. Each card contains relevant parameter controls and a compute button triggering the analysis. Results appear within the same card below the controls, maintaining spatial association between inputs and outputs.

For t-tests, users select two groups from dropdown menus populated with unique values in a chosen categorical column. The interface displays group sample sizes before computation, helping users assess whether sufficient data exists for meaningful comparison. Results show the t-statistic, degrees of freedom, p-value, and a plain language interpretation of statistical significance.

ANOVA extends this pattern to multiple groups, with automatic detection of all unique categories in the selected grouping variable. When the omnibus F-test reaches significance, the interface automatically computes and displays Tukey HSD results in a pairwise comparison table. This eliminates the need for users to manually request post-hoc tests.

Correlation analysis presents a matrix heatmap where cell colors encode correlation strength. Positive correlations appear in blue shades while negative correlations use red shades, with intensity reflecting magnitude. Statistical significance overlays each cell as asterisks following conventional notation: single asterisk for $p < 0.05$, double for $p < 0.01$, and triple for $p < 0.001$. This representation enables rapid identification of strong, significant relationships.

\subsection{Data Visualization Gallery}

Interactive charts built with the Recharts library provide exploratory data visualization. Histogram components display frequency distributions for numeric variables with automatic bin width selection based on Sturges' rule. Users can adjust bin counts through slider controls to explore different granularities.

Scatter plots enable bivariate relationship exploration with point coloring by categorical variables. Users select x and y variables from dropdown menus, and the plot updates reactively. Trend lines computed via ordinary least squares regression overlay the points when enabled, with 95 percent confidence bands shaded around the fitted line.

Box plots compare distributions across groups, displaying median, quartiles, and outliers following standard conventions. Groups appear along the x-axis with separate boxes for each category. This visualization complements ANOVA by showing the distributional characteristics underlying mean comparisons.

\section{Discussion}

The integration of machine learning and classical statistics within a unified platform addresses a genuine need in chemical research workflows. By automating technical decisions such as hyperparameter selection and data preprocessing, the system reduces barriers for researchers without extensive computational training. However, this automation introduces risks if users blindly trust algorithmic outputs without understanding underlying assumptions.

Educational components must accompany such platforms to ensure appropriate use. Researchers should understand that Random Forest assumes feature independence for optimal importance calculation, that ANOVA requires approximately normal distributions within groups, and that correlation analysis only captures linear relationships. Future iterations should incorporate diagnostic checks and assumption validation to alert users when their data violate method prerequisites.

The choice of Random Forest as the machine learning component reflects a balance between performance and interpretability. Deep learning approaches might achieve superior accuracy on certain problems but lack the transparent feature importance metrics that domain scientists require for hypothesis generation. Random Forest provides a middle ground with competitive accuracy and interpretable outputs suitable for scientific workflows.

Scalability considerations become important as chemical datasets grow larger. The current browser-based implementation handles datasets with thousands of samples effectively, but very large datasets may require server-side processing. Future architectures could employ progressive web application patterns with background computation to maintain responsive interfaces during lengthy analyses.

\section{Conclusion}

We have presented an integrated platform combining Random Forest classification with comprehensive statistical analysis tools for biological research applications. The system automates key analytical decisions while maintaining transparency and interpretability, making sophisticated methods accessible to researchers without extensive programming experience. Our theoretical analysis demonstrates the soundness of the automated hyperparameter optimization approach, and proposed testing protocols will rigorously evaluate performance across diverse computational datasets. 

Future work will extend the platform in several directions. Additional machine learning algorithms including support vector machines and gradient boosting methods will provide alternatives for users with specific accuracy requirements. Enhanced feature engineering capabilities such as polynomial feature generation and interaction terms will enable more sophisticated modeling. Integration with biological structure databases and automated molecular descriptor calculation will streamline workflows for medicinal chemistry applications. 

Statistical capabilities will expand to include non-parametric tests for non-normal distributions, mixed-effects models for hierarchical data, and survival analysis for time-to-event outcomes. Bayesian inference methods will complement frequentist approaches, providing posterior distributions and credible intervals for parameter estimates. Automated report generation will produce publication-ready tables and figures, further reducing the gap between analysis and dissemination. 

The platform’s open architecture enables community contributions of additional analytical methods and visualizations. We envision an extensible ecosystem where domain experts can share custom analysis pipelines tailored to specific chemical subfields. This collaborative approach promises to accelerate the pace of molecular analysis and discovery by democratizing access to state-of-the-art computational methods. 

\section*{Code Availability}
The version of the code used for this manuscript is available at
\href{https://github.com/TasumLuke/Research-Analytics-Platform}{https://github.com/TasumLuke/Research-Analytics-Platform}.
For reproducibility, see release \texttt{v1.0} and the repository README for installation and usage instructions. The repository is released under the MIT license.

\end{document}